\documentclass[12pt]{amsart}

\usepackage{amsmath, amssymb, graphics}
\usepackage{graphicx}

\numberwithin{equation}{section}

\newcommand{\mathsym}[1]{{}}

\usepackage{hyperref}
\hypersetup{colorlinks=true,linktoc=page,linkcolor=blue,citecolor=red,urlcolor=blue}

\begin{document}

\title{Worldline approach to noncommutative field theory}
\author{R.\ Bonezzi${}^{^1}$, O.\ Corradini${}^{^{2}}$, S.A.\
  Franchino Vi\~{n}as${}^{^{3,4}}$, P.A.G.\ Pisani${}^{^3}$}
\address{${}^1$ Dipartimento di Fisica, Universit\`a di Bologna
  and\newline INFN, Sezione di Bologna, Via Irnerio 46, I-40126
  Bologna, Italy}
\address{${}^2$ Centro de Estudios en F\'isica y Matem\'aticas
  B\'asicas y Aplicadas\newline Universidad Aut\'onoma de Chiapas,
  Tuxtla Guti\'errez, Mexico}
\address{${}^3$ IFLP-CONICET/ Departamento de F\'isica, Facultad de Ciencias Exactas\newline
Universidad Nacional de La Plata, C.C. 67 (1900), La Plata, Argentina}
\address{${}^4$ Dipartimento di Fisica, Universit\`a di Roma ``La
  Sapienza'' and\newline INFN, Sezione di Roma1, Piazzale Aldo Moro 2,
  I-00185 Roma, Italy\newline}
\email{bonezzi@bo.infn.it, olindo.corradini@unach.mx,\newline safranchino@yahoo.com.ar, pisani@fisica.unlp.edu.ar}
\maketitle

\begin{abstract}
The study of the heat-trace expansion in noncommutative field theory
has shown the existence of Moyal nonlocal Seeley-DeWitt coefficients
which are related to the UV/IR mixing and manifest, in some cases, the
non-renormalizability of the theory. We show that these models can be
studied in a worldline approach implemented in phase space and arrive
at a master formula for the $n$-point contribution to the heat-trace
expansion. This formulation could be useful in understanding some open
problems in this area, as the heat-trace expansion for the
noncommutative torus or the introduction of renormalizing terms in the
action, as well as for generalizations to other nonlocal operators.
\end{abstract}
\maketitle

\section{Introduction}

Noncommutative field theory \cite{Douglas:2001ba,Szabo:2001kg}
involves two concepts which could be considered as fundamental
ingredients of a theory of quantum gravity: nonlocality and a minimal
length scale. As a consequence, the theory presents interesting
renormalization properties, as the so-called UV/IR mixing~\cite{Minwalla:1999px}.

Renormalization in quantum field theory can be implemented with
heat-kernel techniques \cite{Vassilevich:2003xt} since the
one-loop counterterms that regularize the high energy divergencies of
the effective action can be obtained in terms of the Seeley-DeWitt (SDW)
coefficients, that are determined by the asymptotic
expansion, for small values of the proper time, of the heat-trace of a
relevant operator. Such an operator defines the spectrum of
quantum fluctuations and is obtained as the second order correction of
the classical action in a background field expansion.

In a commutative field theory, this operator of quantum fluctuations
is, in general, a (local) differential operator whose SDW coefficients
have been extensively studied; in particular the worldline formalism (WF) has
proved an efficient technique for the computation of these
coefficients~\cite{Bastianelli:2006rx}.

On the other hand, in a noncommutative field theory the operator of
quantum fluctuations is, in general, a nonlocal operator. Lately it has
been shown that the SDW coefficients corresponding to
these types of nonlocal operators have peculiar contributions --which
are related to the non-planar diagrams of the UV/IR mixing-- and are linked to the
 (non)renormalizability of the theory (see the short review~\cite{Vassilevich:2007fq}.)

Let us also mention that in the context of the spectral action
principle~\cite{Chamseddine:1996zu} the SDW coefficients for models defined on
noncommutative spaces have an essential role in the determination of
the corresponding classical actions.

In this article we use WF techniques to obtain a systematic
description of the SDW coefficients of nonlocal operators relevant for
the quantization of noncommutative self-interacting scalar fields on
Moyal Euclidean spacetime. In order to do that we implement the WF in
phase space and we derive a master formula (eq.\ (\ref{mf})) that can
be applied to different settings. We consider this formula to be a
step towards a more systematic understanding of the heat-trace
expansion of nonlocal operators as well as a potentially useful
approach to some open problems in this topic, as those one encounters
for example on the NC torus \cite{Gayral:2006vd}. Further applications to
other nonlocal operators could also be considered.

Although the main motivation of the present article is to develop a new tool for performing one-loop calculations in noncommutative field theories, let us point out that we also apply a path integral representation of quantum mechanical transition amplitudes in Moyal spaces which, as far as we know, has not been used in previous works on noncommutative quantum mechanics. The path integral expressions presented in \cite{Mangano:1997zg,Acatrinei:2001wa} are based on the symplectic form that defines noncommutativity on phase space whereas our worldline formulation depends on the particular representation of the noncommutative algebra given by the Moyal product. Other path integral representations of transition amplitudes have been derived in \cite{Smailagic:2003yb,Gangopadhyay:2009zz} in the coherent states approach to noncommutative quantum mechanics (in \cite{Gangopadhyay:2009zz} coherent states are defined in the Hilbert space consisting on a certain class of Hilbert-Schmidt operators). However, the path integral formulation derived from coherent states is connected instead to the representation of noncommutativity given by the Voros product. There exist applications of path integrals to quantum mechanics in the Moyal plane \cite{Christiansen:2001kv,Dayi:2001af} but in these models the Hamiltonians are quadratic in momenta and coordinates so that, after a Bopp shift, they can be reduced to analogue models in the commutative plane.  Let us mention that the application of path integrals to the Aharonov-Bohm problem in the Moyal plane \cite{Chaichian:2000hy} is performed to leading order in the parameter which defines noncommutativity.

\subsection{Effective action in Euclidean Moyal spacetime}
A field theory on Euclidean Moyal spacetime can be formulated in terms
of the nonlocal Moyal product usually defined as
\begin{equation}\label{moyal}
    (\phi\star \psi)(x):=e^{i\,\partial^{^\phi}\Theta\partial^{^\psi}}\,\phi(x)\psi(x)\,.
\end{equation}
The conditions under which the exponential in this expression is
well-defined are studied in \cite{Estrada:1989da}. The scalar
functions $\phi$ and $\psi$ depend on $x\in\mathbb{R}^d$;
$\partial^{^\phi}$ and $\partial^{^\psi}$ denote their gradients,
respectively. The expression $\partial^{^\phi}\Theta\partial^{^\psi}$
represents
   $\Theta^{ab} \partial^{^\phi}_{x^a} \partial^{^\psi}_{x^b}$,
where $\Theta^{ab}$ are the components of an antisymmetric matrix
$\Theta$ independent of $x$.

With respect to this $\star$ - product, the
coordinates do not commute:
\begin{equation}\label{nc}
    [x^a,x^b]_{\star}:=x^a\star x^b-x^b\star x^a=2i\,\Theta^{ab}\,.
\end{equation}
Thus, the matrix $\Theta$ characterizes the noncommutativity of the
base space. Throughout this article we will consider a possibly
degenerate matrix $\Theta$. Assuming $\Theta$ of rank $2b$ we split
$\mathbb{R}^d$ in a commutative $\mathbb{R}^c$ and a noncommutative
$\mathbb{R}^{2b}$, with $d=c+2b$, by choosing coordinates
$x=(\tilde{x},\hat{x})$ with commuting $\tilde{x}\in\mathbb{R}^{c}$
and noncommuting $\hat{x}\in\mathbb{R}^{2b}$. Consequently, the matrix
$\Theta$ can be written as
\begin{equation}\label{theta}
    \Theta=\mathbf{0}_c\oplus \Xi\,,
\end{equation}
where $\mathbf{0}_c$ is the null matrix in $\mathbb{R}^c$ and $\Xi$ is
a nondegenerate antisymmetric matrix in $\mathbb{R}^{2b}$. In terms of
the Fourier transform
\begin{equation}
    \tilde{\phi}(p)=\int \frac{dx}{(2\pi)^{d}}e^{-ipx}\phi(x)
\end{equation}
the Moyal product reads
\begin{equation}\label{moyal-fourier}
    \widetilde{\phi\star \psi}(p)=\int dq\,e^{-i p\Theta q}\,\tilde{\phi}(p-q)\tilde{\psi}(q)\,.
\end{equation}

A simple model in noncommutative field theory in Euclidean Moyal
spacetime is given by a scalar field with a $\star$ - cubic
self-interaction, described by the Langrangian
\begin{equation}\label{lag}
    \mathcal{L}=\frac{1}{2}(\partial\phi)^2+\frac{m^2}{2}\phi^2+\frac{\lambda}{3!}\phi^3_\star\,,
\end{equation}
where $\phi^3_\star:=\phi\star\phi\star\phi$.

In the ordinary commutative case (i.e.\ $\Theta=0$) the one-loop effective action $\Gamma_{C}$ can be represented as
\begin{equation}
    \Gamma_C=\frac{1}{2}\log{\rm Det}\left\{-\partial^2+m^2+\lambda\,\phi(x)\right\}\,,
\end{equation}
where the Schr\"odinger differential operator between brackets, which
determines the spectrum of quantum fluctuations, arises from the
second functional derivative of the action with respect to the
background field $\phi$. As we have already mentioned, the
regularization of the effective action can be implemented in terms of
the heat-trace of this Schr\"odinger operator.

The spectral theory of Schr\"odinger operators of the type
$-\partial^2+V(x)$ on $\mathbb{R}^d$ shows that for regular potentials
$V(x)$ the heat-trace admits the following asymptotic
expansion in powers of the proper time $\beta>0$ \cite{Gilkey}
\begin{equation}\label{hkc}
    {\rm Tr}\left(f(x)\cdot e^{-\beta
        \left\{-\partial^2+V(x)\right\}}\right)\sim\frac{1}{(4\pi\beta)^{d/2}}\sum_{n=0}^\infty a_n\beta^n\,.
\end{equation}
In this expression we have introduced a smearing function $f(x)$. The
coefficients $a_n$ are called Seeley-DeWitt (SDW) coefficients and are given by
integrals on $\mathbb{R}^d$ of products of the smearing function
$f(x)$, the potential $V(x)$ and derivatives thereof (the reader can find
some proof of this statement in subsection \ref{conmu}.)

In the noncommutative case ($\Theta\neq 0$) the one-loop effective action
$\Gamma_{NC}$ corresponding to the Lagrangian (\ref{lag}) instead results in
\begin{equation}\label{nc-operator}
    \Gamma_{NC}=\frac{1}{2}\log{\rm
      Det}\left\{-\partial^2+m^2+\frac{\lambda}{2}\,L(\phi)+\frac{\lambda}{2}R(\phi)\right\}\,,
\end{equation}
where $L(\phi)$ is an operator whose action on a function $\psi(x)$
is defined as \linebreak $L(\phi)\psi(x):=(\phi\star \psi)(x)$
whereas $R(\phi)\psi(x):=(\psi\star\phi)(x)$. These nonlocal
operators represent the left- respectively right-Moyal multiplication by $\phi$ and can
also be expressed as
\begin{eqnarray}
    L(\phi)\psi(x)=\phi(x+i\Theta \partial)\psi(x)\label{alt1}\,,\\
    R(\phi)\psi(x)=\phi(x-i\Theta \partial)\psi(x)\label{alt2}\,,\nonumber
\end{eqnarray}
where $x\pm i\Theta \partial$ have components $x^a\pm
i\Theta^{ab} \partial_b$.

Thus, in order to regularize the noncommutative effective action
$\Gamma_{NC}$ we must study the heat-trace of the nonlocal operator
\begin{equation}
    -\partial^2+m^2+\frac{\lambda}{2}\,\phi(x+i\Theta\partial)+\frac{\lambda}{2}\,\phi(x-i\Theta\partial)\,.
\end{equation}
Notice that in the case of a quartic self-interaction
$\phi_\star^4$ the operator corresponding to the second functional
derivative of the action contains the following terms:
$L(\phi_\star^2)$, $R(\phi_\star^2)$, $L(\phi)R(\phi)$.

It has been shown that the heat-trace of an operator of the type
\begin{equation}\label{operator}
-\partial^2+L(l_1(x))+R(r_1(x))+L(l_2(x))R(r_2(x))
\end{equation}
also admits an asymptotic expansion in powers of the proper time
$\beta$ \cite{Vassilevich:2005vk} . However, the SDW coefficients are not given, in general, by
integrals on $\mathbb{R}^d$ of local expressions depending on the
smearing function and the potential functions $l_i(x),r_i(x)$.

Nevertheless, if the operator (\ref{operator}) contains only a left-Moyal product (i.e., $r_1(x)=r_2(x)=0$) and the smearing function acts by left-Moyal multiplication, then the SDW
coefficients can be obtained from the commutative ones by replacing
every commutative pointwise product by the noncommutative Moyal
product\footnote{As we will see, to avoid ordering ambiguities, one should consider a commutative operator with matrix-valued coefficients.} \cite{Vassilevich:2003yz,Gayral:2004ww}. The same holds if the
operator contains only a right-Moyal product (i.e., $l_1(x)=l_2(x)=0$)
and the smearing function acts by right-Moyal
multiplication. Therefore, when there is no mixing between left- and right-Moyal multiplications, the SDW coefficients are still integrals of
``local products'' (but in a Moyal sense) of the smearing function,
the potential functions and their derivatives; we will refer to these
as ``Moyal local'' coefficients. In consequence, they are likely to be
introduced as counterterms in the Lagrangian.

On the contrary, for the general case given by expression
(\ref{operator}) some SDW coefficients are not even Moyal local, i.e.\
they cannot be written as integrals of Moyal products of the potential
functions and the smearing function. In some field theories, these
coefficients manifest the non-renormalizability of the corresponding
effective action\footnote{Let us mention that in some noncommutative models in emergent gravity a low-energy regime is to be considered and this UV/IR mixing problem is avoided \cite{Blaschke:2010rr}.} \cite{Gayral:2004cu}.

In this article we develop a worldline approach to the computation of the SDW
coefficients of operator (\ref{operator}). For simplicity, we will
consider the case $l_1(x)=r_1(x)=0$ since the mixing term
$L(l_2)R(r_2)$ will suffice to show the appearance of the (Moyal)
nonlocal coefficients. In order to do that, we write a representation
of the heat-kernel in terms of a path integral in phase space; this is
done in Section \ref{PIphasespace}. This path integral is solved in
Section \ref{GF}, where we compute the corresponding generating
functional. Our main result, the master formula for the noncommutative
heat-trace, is given in Section \ref{nchk}, where we also show how it
works for some particular settings and we confirm existing results
\cite{Vassilevich:2007fq}. In Section \ref{appe} we show a simple application of formula \eqref{mf} to noncommutative field theroy: we consider a scalar field with a quartic self-interaction in Euclidean noncommutative spacetime to show how (Moyal) nonlocal SDW coefficients manifest two different phenomena that could lead to non-renormalizability: the so called UV/IR mixing and the existence of more than one commuting direction in the noncommutative spacetime. Finally, in Section \ref{conclu} we draw our conclusions.

\section{Path Integral in phase space}\label{PIphasespace}

Let us consider an operator which contains a nonlocal potential mixing left- and right-Moyal multiplication
\begin{equation}\label{h}
    H=-\partial^2+L(l)\,R(r)\,,
\end{equation}
where $l(x)$ and $r(x)$ are functions of $x\in\mathbb{R}^d$. Notice
that, due to the associativity of Moyal product, $L(l)$ and
$R(r)$ commute with each other. In accordance with representation
(\ref{alt1}) we write
\begin{equation}\label{ope}
    H=p^2+l(x-\Theta p)\,r(x+\Theta p)\,,
\end{equation}
where $p=-i\partial$. Hamiltonian operator~\eqref{ope} has a fixed ordering given by replacing
coordinate $x$ with operator $x-\Theta p$ in $l$, and with operator $x+\Theta p$ in $r$.
In order to obtain the path integral representation of the transition amplitude associated to~\eqref{ope}
it is convenient to {\it rearrange} the hamiltonian operator in a different form, for example the Weyl ordered form. A phase space operator $A(x,p)$
is written in Weyl-ordered form when it is arranged in such a way that $A(x,p) = A_S(x,p) + \Delta A\equiv A_W(x,p)$, where $A_S(x,p)$
involves symmetric products of $x$'s and $p$'s and $\Delta A$ includes all terms resulting from eventual commutators between $x$'s
and $p$'s, necessary to rearrange $A(x,p)$ in its symmetric form; for
example
$xp  =(xp)_S +\frac12[x,p] = (xp)_S +\frac{i\hbar}{2}\equiv (xp)_W$ (for details see e.g. Appendices B and C in~\cite{Bastianelli:2006rx}).
Above, since  the phase-space operator $l(x-\Theta p)\,r(x+\Theta p)$ mixes coordinates and momenta it is not,
{\it a priori}, written in symmetrized form: in other words  $l(x-\Theta p)\,r(x+\Theta p) = (l(x-\Theta p)\,r(x+\Theta p))_S +\Delta V$.
However, one can show by Taylor expanding the functions $l$ and $r$, that the
operator~\eqref{ope} only involves symmetric products; in other words products of $x$'s and
$p$'s in~\eqref{ope} can be cast in their symmetrized form (for example $(xp)_S
=\frac12 (xp+px)$ and $(x^2 p)_S= \frac13 (x^2 p +xpx +px^2) =
\frac12(x^2 p +p x^2)$) without introducing extra terms, i.e. $\Delta V =0$. Let us, for
example, consider the product of the linear contributions in the
Taylor expansions of $l$ and $r$: it involves the product
$x_-^a x_+^b  \equiv (x^a-\Theta^{ac}p_c)(x^b+\Theta^{bd}p_d)$ whose Weyl-ordering reads
\begin{eqnarray}
  \label{eq:weyl}
x_-^a x_+^b &=& (x^a
  x^b)_S-\Theta^{ac}\Theta^{bd} (p_c p_d)_S -\Theta^{ac} p_c
  x^b+\Theta^{bd} x^a p_d\\
 &=&\nonumber (x_-^a x_+^b)_S -\frac12 \Theta^{ac} [p_c, x^b]+\frac12
 \Theta^{bd} [x^a,p_d] = (x_-^a x_+^b)_S
\end{eqnarray}
thanks to the antisymmetry of the $\Theta^{ab}$ symbol. It is thus easy to
convince oneself that the latter antisymmetry, along with the
commutativity property $[x_-^a,x_+^b]=0$ and the total symmetry of the
coefficients of Taylor expansions of functions $l(x)$ and $r(x)$,
allows to easily prove that all the contributions to the product of
Taylor expansions of operators $l(x-\Theta p)$ and $r(x+\Theta p)$ are
symmetric.

Therefore --using the midpoint
rule~\cite{DeBoer:1995hv,Bastianelli:2006rx}-- one can write the following path integral
representation for the heat-kernel of operator (\ref{ope})
\begin{equation}\label{pi}
\langle x+z\vert e^{-\beta H}\vert x \rangle
        = \int\mathcal{D}x(t)\mathcal{D}p(t)\,e^{-\int_0^{\beta}
          dt\,\left\{p^2(t)-ip(t)\dot{x}(t)\right\}} e^{-\int_0^{\beta}
          dt\,l(x(t)-\Theta p(t))\,r(x(t)+\Theta p(t))}\,,
\end{equation}
where $\beta>0$ and $x(t),p(t)$ represent trajectories in phase space $\mathbb{R}^{2d}$. The path
integral is performed on trajectories $x(t)$ that satisfy the boundary
conditions $x(0)=x$ and $x(\beta)=x+z$ and on trajectories
$p(t)$ that do not satisfy any boundary condition.

It is convenient to replace the integral on the trajectories $x(t)$ by an
integral on perturbations $q(t)$ about the free classical path
$x_{cl}(t)=z~t/\beta+x$. We also make the following rescaling:
$t\rightarrow \beta t$, $q\rightarrow \sqrt{\beta}q$, $p\rightarrow
p/\sqrt{\beta}$. Expression (\ref{pi}) then takes the form
\begin{eqnarray}{\label{pi2}}
        \langle x+z\vert e^{-\beta H}\vert x \rangle
        = \beta^{-d/2}\int\mathcal{D}q\mathcal{D}p\,e^{-\int_0^{1} dt\,\left\{p^2-ip\dot{q}\right\}}\times\\
        \times\,
        e^{i\frac{z}{\sqrt{\beta}}\int_0^1dt\,p}e^{-\beta\int_0^{1}
          dt\,l(x+tz+\sqrt{\beta}q-\Theta p/\sqrt{\beta})
        \,r(x+tz+\sqrt{\beta}q+\Theta p/\sqrt{\beta})}\,,
         \nonumber
\end{eqnarray}
where the perturbations $q$ in configuration space satisfy $q(0)=q(1)=0$.

If we define the mean value of a functional $f[q(t),p(t)]$ as
\begin{equation}
    \left\langle
      f[q(t),p(t)]\right\rangle:=\frac{\int\mathcal{D}q\mathcal{D}p\,e^{-\int_0^{1}
      dt\,\left\{p^2-ip\dot{q}\right\}} f[q(t),p(t)]}{\int\mathcal{D}q\mathcal{D}p\,e^{-\int_0^{1}
      dt\,\left\{p^2-ip\dot{q}\right\}}}
\end{equation}
then the transition amplitude (\ref{pi2}) reads
\begin{equation}
        \langle x+z\vert e^{-\beta H}\vert x \rangle
        = \frac1{(4\pi\beta)^{d/2}}\left\langle
          e^{i\frac{z}{\sqrt{\beta}}\int_0^1dt\,p}e^{-\beta\int_0^{1}
            dt\,l(x+tz+\sqrt{\beta}q-\Theta p/\sqrt{\beta})
        \,r(x+tz+\sqrt{\beta}q+\Theta p/\sqrt{\beta})}\right\rangle\,.
\end{equation}
Next, we make a small $\beta$ expansion of the second exponential and
a Taylor expansion of $l$ and $r$ around $x$
\begin{eqnarray}{\label{pi3}}
        \langle x+z\vert e^{-\beta H}\vert x \rangle
        =\frac1{(4\pi\beta)^{d/2}}\sum_{n=0}^\infty \frac{(-\beta)^n}{n!}\int_0^1dt_1\ldots\int_0^1dt_n\times\\\nonumber\times\,
        \left\langle e^{i\frac{z}{\sqrt{\beta}}\int_0^1dt\,p}
        e^{\sum_{i=1}^n[t_iz+\sqrt{\beta}q(t_i)-\Theta p(t_i)/\sqrt{\beta}]\partial^l_i+
        [t_iz+\sqrt{\beta}q(t_i)+\Theta
        p(t_i)/\sqrt{\beta}]\partial^r_i}\right\rangle\times\\\nonumber\left.\times\,
        l(x_1)\ldots l(x_n)\,r(x_1)\ldots r(x_n)\right|_{x}\,,
\end{eqnarray}
where $\partial^l_i,\partial^r_i$ are the gradients\footnote{Notice
  that the subindex $i$ does not denote spacetime components but refers to the $i$-th spacetime point $x_i$.} of
$l(x_i),r(x_i)$. As indicated, all $x_i$ must be evaluated at $x$
after performing these derivatives. Expression (\ref{pi3}) can be
written as
\begin{eqnarray}{\label{pi4}}
        \langle x+z\vert e^{-\beta H}\vert x \rangle
        =\frac1{(4\pi\beta)^{d/2}}\sum_{n=0}^\infty \frac{(-\beta)^n}{n!}\int_0^1dt_1\ldots\int_0^1dt_n\times\\\nonumber\left.\times\,
        e^{\sum_{i=1}^n t_iz(\partial^l_i+\partial^r_i)}\left\langle
        e^{\int_0^1 dt \left(p\, k_n +q\, j_n\right)}
        \right\rangle
        l(x_1)\ldots l(x_n)\,r(x_1)\ldots r(x_n)\right|_{x}
         \nonumber
\end{eqnarray}
if we define the sources $k_n(t),j_n(t)$ as
\begin{eqnarray}
k_n(t)=\frac{iz}{\sqrt{\beta}}+\frac{\Theta}{\sqrt{\beta}}\sum_{i=1}^{n}\delta(t-t_i)(\partial^l_i-
\partial^r_i)\,,
\label{k}\\
j_n(t)=\sqrt{\beta}\,\sum_{i=1}^{n}\delta(t-t_i)(\partial^l_i+\partial^r_i)\label{j}\,.\nonumber
\end{eqnarray}
In the following section we will compute the expectation value
$\left\langle e^{\int_0^1 dt
    \left(p\, k +q\, j\right)}\right\rangle$ for arbitrary
sources $k(t),j(t)$.

\section{The generating functional in phase space}\label{GF}

Let us compute the generating functional
\begin{eqnarray}
Z[k,j]:=\left\langle e^{\int_0^1 dt \left(p\, k +q\, j \right)}\right\rangle 
=\frac{\int\mathcal{D}q\mathcal{D}p\,e^{-\int_0^{1} dt\,\left(p^2-ip\,\dot{q}\right)}
e^{\int_0^{1} dt\,(p\, k +q\, j)}}{\int\mathcal{D}q\mathcal{D}p\,e^{-\int_0^{1} dt\,\left(p^2-ip\,\dot{q}\right)}}\\[1mm]\nonumber
=\frac{\int\mathcal{D}P\ e^{-\frac{1}{2}\int_0^{1} dt\,P^t A P
+\int_0^1dt\,P^tK}}{\int\mathcal{D}P\ e^{-\int_0^{1} dt\,P^t A P}}
\end{eqnarray}
for arbitrary sources $k(t), j(t)$. In this last expression we have defined the vectors
\begin{eqnarray}
    P:=\left(\begin{array}{c}p(t)\\q(t)\end{array}\right)&
    K:=\left(\begin{array}{c}k(t)\\j(t)\end{array}\right)
\end{eqnarray}
and the operator
\begin{equation}\label{propagator}
A:=
\begin{pmatrix}
2&-i\partial_t\\
i\partial_t&0
\end{pmatrix}
\end{equation}

Completing squares and inverting the operator $A$ --taking into
account the boundary conditions $q(0)=q(1)=0$-- we obtain the
generating functional in phase space
\begin{equation}\label{z}
Z[k,j]=e^{\frac{1}{2}\int_0^1 dt\, K^t A^{-1} K}\,.
\end{equation}
The kernel of the operator $A^{-1}$ is given by
\begin{equation}
A^{-1}(t,t')=
\begin{pmatrix}
\frac{1}{2}&\frac{i}{2}\left[h(t,t')+f(t,t')\right]\\
\frac{i}{2}\left[h(t,t')-f(t,t')\right]&2g(t,t')
\end{pmatrix}\,,
\end{equation}
where
\begin{eqnarray}
h(t,t'):=1-t-t'\,,\\
f(t,t'):=t-t'-\epsilon(t-t')\,,\nonumber\\
g(t,t'):=t(1-t')H(t'-t)+t'(1-t)H(t-t')\,.\nonumber
\end{eqnarray}
The sign function $\epsilon(\cdot)$ is $\pm 1$ if its argument is
positive or negative, respectively; $H(\cdot)$ represents the
Heaviside function.

To obtain the expectation value of expression (\ref{pi4}) we replace
the sources given by eqs.\ (\ref{k}) into expression
(\ref{z})
\begin{equation}\label{ev}
    \left\langle e^{\int_0^1 dt \left(p\, k_n +q\, j_n \right)}\right\rangle=
    e^{-\frac{z^2}{4\beta}+\frac{iz}{2\beta}\Theta\sum_{i=1}^n(\partial^l_i-\partial^r_i)}
    e^{\triangle_n}\,,
\end{equation}
where the operator $\triangle_n$ is defined as
\begin{eqnarray}\label{dn}\\\nonumber
    \triangle_n:=\sum_{i,j=1}^n\left[\beta
      g(t_i,t_j)(\partial^l_i+\partial^r_i)(\partial_j^l+\partial_j^r)-
      \frac{1}{4\beta}(\partial^l_i-\partial^r_i)\Theta^2(\partial_j^l-\partial_j^r)\right.
      \\\nonumber\left.\mbox{}
      -\frac{i}{2}f(t_i,t_j)(\partial_i^l\Theta\partial_j^l-\partial_i^r\Theta\partial_j^r)
      -ih(t_i,t_j)\partial_i^l\Theta\partial_j^r\right]\,.
\end{eqnarray}

\section{The noncommutative heat-kernel}\label{nchk}

The smeared heat-trace of the nonlocal operator $H$, defined in eq.\ (\ref{ope}), can be written as
\begin{equation}\label{tr}
    {\rm Tr}\left(f(x)\,\bar{\star}\, e^{-\beta H}\right)=\int_{\mathbb{R}^d}
    dx\,\left.\langle x+z\vert e^{-\beta H}\vert x \rangle\right|_{z=-i\bar{\Theta}\partial^f}f(x)\,,
\end{equation}
where $\bar{\star}$ represents a Moyal product as defined in (\ref{moyal}) but in terms of
another antisymmetric matrix $\bar{\Theta}$ which, in principle,
differs from $\Theta$. Expression~\eqref{tr} can be easily obtained by inserting in the l.h.s a spectral decomposition of unity in terms of position eigenstates and using that $e^{-i\bar\Theta \partial^{f}\partial} \langle x| =\langle x-i\bar\Theta \partial^{f} |$.

The matrix $\bar{\Theta}$ allows us to consider at the same time the
cases where the smearing function $f$ acts by commutative pointwise
multiplication ($\bar{\Theta}=0$), by left-Moyal multiplication
($\bar{\Theta}=\Theta$) or right-Moyal multiplication
($\bar{\Theta}=-\Theta$). As indicated, the variable
$z$ must be replaced by the operator $-i\bar{\Theta}\partial^f$, where
$\partial^f$ is the gradient acting on the smearing function $f$
only. The heat-kernel $\langle x+z\vert e^{-\beta H}\vert x \rangle$
is obtained by replacing eq.\ (\ref{ev}) into expression~(\ref{pi4}).
Inserting the result into~(\ref{tr}) we get our master
formula
\begin{eqnarray}{\label{mf}}
        {\rm Tr}\left(f(x)\,\bar{\star}\, e^{-\beta H}\right)
        =\frac{1}{(4\pi\beta)^{d/2}}\sum_{n=0}^\infty (-\beta)^n
        \int_{\mathbb{R}^d}dx\,f(x)\,e^{\frac{1}{4\beta}\sum_{i,j=1}^n D_iD_j}
        \times\\
        \nonumber\left.\times\,
        \int_0^1 dt_1\int_0^{t_1}dt_2\ldots\int_0^{t_{n-1}}dt_n
        \ e^{i\triangle^{NC}_n+\beta\triangle^{C}_n}
        l(x_1)\ldots l(x_n)\,r(x_1)\ldots r(x_n)
        \right|_{x}\,,
\end{eqnarray}
where we have defined the following differential operators
\begin{eqnarray}\label{df}
    \triangle^{C}_n:=\sum_{i,j=1}^n g(t_i,t_j)(\partial^l_i+\partial^r_i)(\partial^l_j+\partial^r_j)\,,
    \\[1mm]
    D_i:=\left(\Theta-\bar{\Theta}\right)\partial^l_i-\left(\Theta+\bar{\Theta}\right)\partial^r_i\,,
\nonumber
    \\[1mm]
    \triangle^{NC}_n:=
    \sum_{i< j=1}^n\Big\{\left[\partial^l_i\Theta\partial^l_j-\partial^r_i\Theta\partial^r_j\right]
     -(1-t_i-t_j)(\partial^l_i\Theta\partial^r_j-\partial^r_i\Theta\partial^l_j)
    \nonumber\\
    \mbox{} + (t_i-t_j)\left[\partial^l_i(-\Theta+\bar{\Theta})\partial^l_j+\partial^r_i(\Theta+\bar{\Theta})
\partial^r_j+
\partial^l_i\bar\Theta\partial^r_j+\partial^r_i\bar\Theta\partial^l_j\right]\Big\}
    \nonumber\\
    -\sum_{i=1}^n (1-2t_i)\partial^l_i\Theta\partial^r_i\,.\nonumber
\end{eqnarray}
In the derivation of (\ref{mf}) we have used the symmetry of the
integrand with respect to permutations of the variables $t_i$ and we
have integrated by parts to make the replacement
$\partial^f\rightarrow-\sum_{i=1}^n(\partial^l_i+\partial^r_i)$.

A few remarks are now in order. As we will see next, the SDW
coefficients for the commutative case are fully determined by the
action of the operator $\triangle_n^C$ since $\triangle_n^{NC}$ and
$D_i$ vanish for $\Theta=\bar\Theta=0$. On the other hand, if
$r(x)\equiv 1$ (or $l(x)\equiv 1$) and the smearing function acts by
left- (respectively right-) Moyal multiplication, then $D_i$ vanishes
and the only non-vanishing term in $\triangle_n^{NC}$ is the first one
$\partial^l_i\Theta\partial^l_j$ (respectively
$-\partial^r_i\Theta\partial^r_j$) which replaces any pointwise
product by a left- (respectively right-) Moyal product.  Let us also mention that the
heat-trace with no smearing function corresponds to $f(x)\equiv 1$ and
$\bar\Theta=0$.

Finally, we will show that the $1/\beta$ coefficient of $\sum_{i,j} D_iD_j$ in expression \eqref{mf} is responsible for Moyal nonlocal SDW coefficients, which can be shown to correspond to contributions of non-planar diagrams, leading to the well-known UV/IR mixing.

In the rest of this section, we will apply master formula (\ref{mf})
to these different settings in order to describe the SDW
coefficients.

\subsection{Commutative limit}\label{conmu}

First of all, we apply formula (\ref{mf}) to the heat-trace in the
commutative case for future comparison with the subsequent
noncommutative expressions. For $\bar{\Theta}=\Theta=0$ it is
sufficient to consider the case $r(x)\equiv 1$. As already mentioned,
the operators $D_i$ and $\triangle_n^{NC}$ vanish so that formula
(\ref{mf}) reads
\begin{eqnarray}\label{cc}
        {\rm Tr}\left(f(x)\,\cdot\, e^{-\beta H}\right)
        =\frac{1}{(4\pi\beta)^{d/2}}\sum_{n=0}^\infty (-\beta)^n
        \int_{\mathbb{R}^d}dx\,f(x)
        \times\\
        \nonumber\left.\times\,
        \int_0^1 dt_1\int_0^{t_1}dt_2\ldots\int_0^{t_{n-1}}dt_n
        \ e^{\beta\sum_{i,j=1}^n g(t_i,t_j)\partial_i\partial_j}
        \ l(x_1)\ldots l(x_n)
        \right|_{x}\,.
\end{eqnarray}
This expression shows that the SDW coefficients $a_n$ are integrals of
products of the smearing function, the potential and derivatives
 thereof (see eq.~\eqref{hkc}.)

For later use, in expression~\eqref{cc}  we keep the time-ordering explicit in the multi worldline integral so that~\eqref{cc} gives the correct SDW coefficients even if the potential $l(x)$ is a matrix potential. In such a case the product of two adjacent $l(x)$'s has to be understood as a spatially-pointwise matricial product.

\subsection{Pointwise multiplication by a smearing function}\label{tz}

In this subsection we consider the noncommutative operator (\ref{ope})
but for the case in which the smearing function $f(x)$ acts by
pointwise multiplication, i.e.\ $\bar{\Theta}=0$. We will show that
some SDW coefficients are Moyal nonlocal  \cite{Vassilevich:2003yz}.

For $\bar{\Theta}=0$ formula (\ref{mf}) reads
\begin{eqnarray}{\label{tbz}}
        {\rm Tr}\left(f(x)\,\cdot\, e^{-\beta H}\right)
        =\frac{1}{(4\pi\beta)^{d/2}}\sum_{n=0}^\infty (-\beta)^n
        \int_{\mathbb{R}^d}dx\,f(x)\,
        \times\\
        \nonumber\left.\times\,
        \int_0^1 dt_1\int_0^{t_1}dt_2\ldots\int_0^{t_{n-1}}dt_n
        \ e^{\triangle_n}
        l(x_1)\ldots l(x_n)\,r(x_1)\ldots r(x_n)
        \right|_{x}\,.
\end{eqnarray}
For the purposes of this section it will suffice to consider a
potential which contains only a left-Moyal product so we will
consider the case $r(x)\equiv 1$. Thus, the leading
terms in (\ref{tbz}) read
\begin{equation}\label{pf}
        {\rm Tr}\left(f(x)\,\cdot\, e^{-\beta H}\right)
        =\frac{1}{(4\pi\beta)^{d/2}}
        \int_{\mathbb{R}^d}dx\,f(x)\left(1-\beta\,
        e^{-\frac{1}{4\beta}\partial^l\Theta^2\partial^l}
        l(x)+\ldots\right)\,.
\end{equation}
The first term coincides with the leading term --the volume
contribution-- in the commutative case, whereas the second one can be
written as (see the discussion below eq.\ (\ref{nc}))
\begin{eqnarray}
\\\nonumber
        -\frac{\beta}{(4\pi\beta)^{d/2}}
        \int_{\mathbb{R}^c}d\tilde{x}\int_{\mathbb{R}^{2b}}d\hat{x}\,f(\tilde{x},\hat{x})\int_{\mathbb{R}^{2b}}
        d\hat{y}
        \frac{\beta^{b}}{\pi^{b}\ {\rm det}\,\Xi}
        \ e^{\beta(\hat{x}-\hat{y})\Xi^{-2}(\hat{x}-\hat{y})}
        \ l(\tilde{x},\hat{y})\sim\\\nonumber
        \sim -\frac{1}{(4\pi\beta)^{d/2}}\cdot\frac{\beta^{b+1}}{\pi^b\ {\rm det}\,\Xi}
        \int_{\mathbb{R}^c}d\tilde{x}
        \left(\int_{\mathbb{R}^{2b}}d\hat{x}\,f(\tilde{x},\hat{x})\right)
        \left(\int_{\mathbb{R}^{2b}}d\hat{y}\,l(\tilde{x},\hat{y})\right)+\ldots
\end{eqnarray}
As can be seen from this expression, the presence of a smearing
function that acts by pointwise multiplication originates a contribution to the SDW coefficient $a_{b+1}$
which is nonlocal in the generalized Moyal sense. Notice that this
Moyal nonlocal coefficient was obtained when considering the case
$r(x)\equiv 1$, i.e. even when the
potential in $H$ does not mix left- and right-Moyal actions.

\subsection{No mixing between left- and right-Moyal multiplication}

We will now consider the case in which the heat-trace involves only
left-Moyal multiplication ($r(x)\equiv 1$ and $\bar\Theta=\Theta$) or
only right-Moyal multiplication ($l(x)\equiv 1$ and
$\bar\Theta=-\Theta$). Under any of these alternative assumptions the
operators $D_i$ vanish, whereas $\triangle_n^{NC}$ takes the form
\begin{equation}\label{dfolr}
    \triangle^{NC}_n:=
    \pm\sum_{i< j=1}^n\partial_i\Theta\partial_j\,,
\end{equation}
where the upper (lower) sign corresponds to the case where only left-
(right-) Moyal multiplication is considered; the derivatives
$\partial_i$ act consequently on $l(x_i)$ ($r(x_i)$). Therefore,
according to formula (\ref{mf}), the action of the operator
$e^{i\triangle_{n}^{NC}}$ is the only effect of noncommutativity on
the SDW coefficients. Notice that the action of these operator defines
Moyal multiplication (see eq.\ (\ref{moyal})), so that the heat-trace (for the left-Moyal case) reduces to
\begin{eqnarray}\label{ncol}
        {\rm Tr}\left(f(x)\star e^{-\beta H}\right)
        =\frac{1}{(4\pi\beta)^{d/2}}\sum_{n=0}^\infty (-\beta)^n
        \int_{\mathbb{R}^d}dx\,f(x)
        \times\\
        \nonumber\left.\times\,
        \int_0^1 dt_1\int_0^{t_1}dt_2\ldots\int_0^{t_{n-1}}dt_n
        \ e^{\beta\sum_{i,j=1}^n g(t_i,t_j)\partial_i\partial_j}
        \ l(x_1)\star \cdots \star l(x_n)
        \right|_{x}\,.
\end{eqnarray}
In this case, we conclude that the noncommutative SDW coefficients can be obtained from the commutative coefficients for a matrix potential $l(x)$ (cfr.~\eqref{cc}), by replacing every spatially-pointwise matrix product of adjacent potentials with left- (right-) Moyal products. For example, up to $\beta^3$ one gets
\begin{eqnarray}
 {\rm Tr}\left(f(x)\star e^{-\beta H}\right)
        =\frac{1}{(4\pi\beta)^{d/2}} \int_{\mathbb{R}^d}dx\,f(x) \times \\ \nonumber \times \Biggl[1-\beta l(x) +\beta^2 \left( \frac12 l_\star^2(x) -\frac16\partial^2 l(x)\right)+\beta^3 \Biggl( -\frac1{60}\partial^4 l(x)\\ \nonumber  +\frac1{12} \Big(\partial^2l \star l(x) +l\star \partial^2 l(x) +\partial l\star \partial l(x)\Big)-\frac1{3!} l_\star^3(x)\Biggr) 
        +\cdots\Biggr]
\end{eqnarray}
In summary, when there is only left- (right-) Moyal multiplication the SDW coefficients are Moyal local~\cite{Vassilevich:2003yz,Gayral:2004ww}.

\subsection{Mixing between left- and right-Moyal multiplication}

In this last subsection we will consider the presence of both functions $l(x)$
and $r(x)$ in the potential of the operator (\ref{ope}). The
calculation goes along the same line followed in subsection
\ref{tz} and we will arrive to a similar conclusion, namely, the
existence of Moyal nonlocal SDW coefficients. For
simplicity, we will consider $f(x)\equiv 1$. Formula (\ref{mf}) then reads
\begin{eqnarray}\label{lrm}
        {\rm Tr}\left(e^{-\beta H}\right)
        =\frac{1}{(4\pi\beta)^{d/2}}\sum_{n=0}^\infty (-\beta)^n
        \int_{\mathbb{R}^d}dx\,
        \times\\
        \nonumber\left.\times\,
        \int_0^1 dt_1\int_0^{t_1}dt_2\ldots\int_0^{t_{n-1}}dt_n
        \ e^{\triangle_n}
        l(x_1)\ldots l(x_n)\,r(x_1)\ldots r(x_n)
        \right|_{x}\,,
\end{eqnarray}
Let us consider the leading terms in expansion (\ref{lrm})
corresponding to an increasing number $n$ of insertions of $l(x)$ and
$r(x)$.

For $n=0$ we get the commutative volume contribution. For $n=1$ we get the leading contribution
\begin{eqnarray}\label{n=1}
        -\frac{\beta}{(4\pi\beta)^{d/2}}
        \int_{\mathbb{R}^d}dx\,
        r(x)e^{-\frac{1}{\beta}\partial^l\Theta^2\partial^l}l(x)\,,
\end{eqnarray}
where upon integration by parts we have replaced
$\partial^r\rightarrow-\partial^l$. Expression (\ref{n=1}) can be obtained from
the subleading term of (\ref{pf}) by identifying $r(x)$ with
$f(x)$. Proceeding analogously we get for (\ref{n=1})
\begin{eqnarray}\label{nlc}
        \mbox{}\\\nonumber
        -\frac{1}{(4\pi\beta)^{d/2}}\frac{\beta^{b+1}}{(4\pi)^{b}\ {\rm det}\,\Xi}
        \int_{\mathbb{R}^c}d\tilde{x}\int_{\mathbb{R}^{2b}}d\hat{x}\,r(\tilde{x},\hat{x})
        \int_{\mathbb{R}^{2b}}d\hat{y}\ e^{\frac{\beta}{4}(\hat{x}-\hat{y})\Xi^{-2}(\hat{x}-\hat{y})}
        \ l(\tilde{x},\hat{y})\sim\\\nonumber
        \sim -\frac{1}{(4\pi\beta)^{d/2}}\cdot\frac{\beta^{b+1}}{(4\pi)^b\ {\rm det}\,\Xi}
        \int_{\mathbb{R}^c}d\tilde{x}
        \left\{
        \left(\int_{\mathbb{R}^{2b}}d\hat{x}\,r(\tilde{x},\hat{x})\right)
        \left(\int_{\mathbb{R}^{2b}}d\hat{y}\,l(\tilde{x},\hat{y})\right)\right.\\\nonumber
        \mbox{}+\frac{\beta}{4}(\Xi^{-2})_{ij}
        \left[
        \int_{\mathbb{R}^{2b}}d\hat{x}\,\hat{x}^i\hat{x}^jr(\tilde{x},\hat{x})
        \cdot
        \int_{\mathbb{R}^{2b}}d\hat{y}\,l(\tilde{x},\hat{y})
        \right. 
        + \int_{\mathbb{R}^{2b}}d\hat{x}\,r(\tilde{x},\hat{x})
        \cdot
        \int_{\mathbb{R}^{2b}}d\hat{y}\,\hat{y}^i\hat{y}^jl(\tilde{x},\hat{y})
        \\\nonumber
        \left.\left.
        \mbox{}-2\int_{\mathbb{R}^{2b}}d\hat{x}\,\hat{x}^ir(\tilde{x},\hat{x})
        \int_{\mathbb{R}^{2b}}d\hat{y}\,\hat{y}^jl(\tilde{x},\hat{y})
        \right]
        +\ldots
        \right\}\,.
\end{eqnarray}
In this expression we have explicitly written the splitting of the
coordinates $x\in\mathbb{R}^d$ in commuting $\tilde{x}\in\mathbb{R}^c$
and noncommuting $\hat{x}\in\mathbb{R}^{2b}$ coordinates (see the
discussion below eq.\ (\ref{nc})). Expansion (\ref{nlc}) shows the
Moyal nonlocal contributions, which are linear in the product
$r(x)l(y)$, to the coefficients $a_{b+1}$ and $a_{b+2}$
\cite{Vassilevich:2005vk} that could lead to the non-renormalizability
of the corresponding theory \cite{Gayral:2004cu}.

\subsection{The noncommutative torus}\label{nctorus}

As our last example, we will consider the operator (\ref{ope}) on the
$d$-dimensional noncommutative torus $T^d$, as defined in \cite{Gayral:2006vd}. The coordinates $x=(\tilde{x},\hat{x})$ on $T^d$ can be split in commuting $\tilde{x}\in T^{c}$
and noncommuting $\hat{x}\in T^{2b}$ components, with $d=c+2b$, as discussed after eq.\ (\ref{nc}). We also define $0\leq x_i\leq L_i$.

The heat-kernel $\langle y|e^{-\beta
  H}|x\rangle_{T^d}$ subject to the corresponding periodic boundary
conditions can be obtained as the sum of infinitely many transition
amplitudes $\langle y+t_k|e^{-\beta H}|x\rangle_{\mathbb{R}^d}$ computed
on the whole space $\mathbb{R}^d$, where $t_k=(L_1 k_1,\ldots,L_d
k_d)$ and $k=(k_1,\ldots,k_d)\in\mathbb{Z}^d$:
\begin{equation}
    \langle y|e^{-\beta H}|x\rangle_{T^d}=\sum_{k\in \mathbb{Z}^d}\langle y+t_k|e^{-\beta H}|x\rangle_{\mathbb{R}^d}
\end{equation}
The functions $l,r$ in the operator $H$ of the r.h.s.\ of this equation are the periodic
extensions to the whole space $\mathbb{R}^d$ of the functions $l,r$ in the operator $H$ of the l.h.s.

Correspondingly, the heat-trace on the torus can be computed as
\begin{equation}
    {\rm Tr}\left(e^{-\beta H}\right)=\int_{T^d}dx\, \langle x|e^{-\beta H}|x\rangle_{T^d}=
    \int_{T^d}dx\,\sum_{k\in \mathbb{Z}^d}\langle x+t_k|e^{-\beta H}|x\rangle_{\mathbb{R}^d}
\end{equation}
The transition amplitudes on the whole $\mathbb{R}^d$ were already computed and can be
read from eqs.\ (\ref{pi4}) and (\ref{ev}). The first contribution, corresponding to $n=0$ in eq.\ (\ref{pi4}), gives the volume term
\begin{equation}
    \frac{1}{(4\pi\beta)^{d/2}}\,\prod_{i=1}^d L_i\sum_{k\in \mathbb{Z}}
    e^{-\frac{L_i^2}{4\beta} k^2}\sim \frac{1}{(4\pi\beta)^{d/2}}\,L_1\ldots L_d\,.
\end{equation}
Notice that the only contribution to the asymptotic expansion for small $\beta$ comes from the term corresponding to $k=0$.

On the other hand, the term corresponding to $n=1$ in eq.\ (\ref{pi4}) reads
\begin{equation}\label{n1}
    -\frac{\beta}{(4\pi\beta)^{d/2}}\sum_{k\in \mathbb{Z}^d}\int_{T^d}dx\,r(x)
    e^{-\frac{1}{\beta}\left(-i\Theta\partial+\frac{1}{2}t_k\right)^2}
    l(x)\,,
\end{equation}
where we have used the periodicity of $l(x)$ and $r(x)$ to replace $\partial^r$ by $-\partial^l$. This contribution has been analyzed in \cite{Gayral:2006vd} for the case in which the matrix elements of $\Theta$ satisfy a Diophantine condition.

\section{Application to $\lambda\, \phi\star^4$}\label{appe}

In this section we will apply our formula \eqref{mf} to a simple noncommutative model. In particular, we will compute some SDW coefficients to study the one-loop corrections to the propagator of a real scalar field with a quartic self-interaction defined on the Euclidean spacetime $\mathbb{R}^d$. We therefore consider the following action
\begin{equation}
    \mathcal{L}=\frac{1}{2}(\partial\phi)^2+\frac{m^2}{2}\phi^2+\frac{\lambda}{4!}\phi^4_\star\,,
\end{equation}
where $\phi^4_\star:=\phi\star\phi\star\phi\star\phi$. The one-loop effective action $\Gamma$ can be written as \cite{Vassilevich:2005vk}
\begin{eqnarray}\label{effactapp}
    \Gamma=\frac{1}{2}\log{\rm Det}\left\{-\partial^2+m^2+\frac{\lambda}{3}\,[L(\phi_\star^2)+R(\phi_\star^2)
    +L(\phi)R(\phi)]\right\}\\ \nonumber
    =-\frac{1}{2}\int_{\Lambda^{-2}}^\infty \frac{d\beta}{\beta}\,e^{-\beta m^2}\,{\rm Tr}\,e^{-\beta
    \left\{-\partial^2+\frac{\lambda}{3}\,\left[ L(\phi_\star^2)+R(\phi_\star^2)
    +L(\phi)R(\phi)\right] \right\}}\,,
\end{eqnarray}
where we have used the Schwinger proper time approach to represent the functional determinant and we have introduced an UV-cutoff $\Lambda$.

Since the propagator is obtained from the quadratic terms in the effective action, we only need to consider the terms in expression \eqref{effactapp} which are linear in $\lambda$. Therefore, we can use formula $\eqref{mf}$ to compute the contributions of the three terms $L(\phi_\star^2)$, $R(\phi_\star^2)$ and $L(\phi)R(\phi)$ separately. By replacing $f(x)\equiv 1$, $\bar\Theta=0$ and $r(x)\equiv 1$ in the term corresponding to $n=1$ in eq.\ \eqref{mf} we obtain the contribution of $L(\phi_\star^2)$ to the effective action. Analogously, the contribution corresponding to $R(\phi_\star^2)$ is obtained by replacing instead $l(x)\equiv 1$. Both contributions are equal and the sum of them reads
\begin{equation}\label{propapp}
  \int_{\Lambda^{-2}}^\infty d\beta\,e^{-\beta m^2}\,\frac{1}{(4\pi \beta)^{d/2}}
  \int_{\mathbb{R}^d}dx\,\frac{\lambda}{3}\phi_\star^2(x)=
  \frac{\lambda}{3}\frac{m^{d-2}}{(2\pi)^{d/2}}\,\Gamma(1-d/2,m^2/\Lambda^2)\int_{\mathbb{R}^d}\phi^2\,,
\end{equation}
where $\Gamma(\cdot,\cdot)$ represents the incomplete gamma function \cite{A-S}. In the limit $\Lambda\rightarrow\infty$, this contribution diverges as $\log{\Lambda}$, for $d=2$, and as $\Lambda^{d-2}$, for $d>2$. This divergence is eliminated by a mass redefinition; in this way, the dependence of the mass with the cutoff $\Lambda$ is determined. Notice that result \eqref{propapp} --which corresponds to the contribution of one-loop planar diagrams-- does not depend on the noncommutativity parameters and holds also in the commutative case.

The remaining contribution, corresponding to the term $L(\phi)R(\phi)$, is obtained by replacing $f(x)\equiv 1$, $\bar\Theta=0$ and $\partial^l=-\partial^r$ in the term corresponding to $n=1$ in eq.\ \eqref{mf}; the result reads,
\begin{eqnarray}\label{propappmixed}
  \frac{1}{2}\int_{\Lambda^{-2}}^\infty d\beta\,e^{-\beta m^2}\,\frac{1}{(4\pi \beta)^{d/2}}
  \int_{\mathbb{R}^d}dx\,\frac{\lambda}{3}\phi(x)e^{-\frac{1}{\beta}\partial\Theta^2\partial}\phi(x)=\\\nonumber
  =\frac{\lambda}{6}(2\pi)^{3d/2}m^{d-2}\,\int d^c\tilde{p}\,d^{2b}\hat{p}
  \ \tilde\phi^*(\tilde{p},\hat{p})\tilde\phi(\tilde{p},\hat{p})
  \cdot\Sigma_{NP}(\hat{p})
  \,,
\end{eqnarray}
where
\begin{equation}\label{bess}
  \Sigma_{NP}(\hat{p}):=\int_0^\infty \frac{d\beta}{\beta^{d/2}}\,e^{-\beta-\frac{m^2}{\beta}|\Xi \hat{p}|^2}=
  2(m|\Xi \hat{p}|)^{1-d/2}K_{d/2-1}(2m|\Xi \hat{p}|)\,,
\end{equation}
being $K_{d/2-1}(\cdot)$ the modified Bessel function. In eq.\ \eqref{propappmixed} we have made use of the splitting defined in eq.\ \eqref{theta} to separate spacetime $\mathbb{R}^d$ into $\mathbb{R}^c$, which is described by commuting coordinates, and $\mathbb{R}^{2b}$, where noncommutativity is defined by the non-degenerate matrix $\Xi$. We have also written this contribution to the effective action in terms of the Fourier transform $\tilde\phi(\tilde{p},\hat{p})$ of the field, where $\tilde{p}\in\mathbb{R}^c$ and $\hat{p}\in\mathbb{R}^{2b}$. Notice also that in expression \eqref{bess} the term depending on $\Xi$ makes the integral convergent at $\beta\rightarrow 0$ and, in consequence, the cutoff $\Lambda$ has been removed. In other words, noncommutativity regularizes UV-divergence at the one-loop level. It can be shown that expression \eqref{propappmixed} corresponds to the contribution of one-loop non-planar diagrams.

Since $\Sigma_{NP}(\hat{p})\sim |\hat{p}|^{-d+2}$, for small $|\hat{p}|$ and $d>2$, the integrand in \eqref{propappmixed} grows as $|\hat{p}|^{1-c}$ for $|\hat{p}|\rightarrow 0$, where $c=d-2b$ is the number of commuting coordinates. Therefore, this contribution to the one-loop effective action is divergent if the number of commuting coordinates is greater or equal than two. The same result can be obtained from expression \eqref{nlc}, whose leading term reads
\begin{equation}\label{hnl}
        -\frac{1}{(4\pi\beta)^{d/2}}\cdot\frac{\beta^{b+1}}{(4\pi)^b\ {\rm det}\,\Xi}
        \int_{\mathbb{R}^c}d\tilde{x}
        \left(\int_{\mathbb{R}^{2b}}d\hat{x}\,\phi(\tilde{x},\hat{x})\right)
        \left(\int_{\mathbb{R}^{2b}}d\hat{y}\,\phi(\tilde{x},\hat{y})\right)\,.
\end{equation}
If this contribution is inserted in the last line of expression \eqref{effactapp} one can see that the integrand behaves as $\beta^{-c/2}$ for small $\beta$ and then the integral diverges as $\Lambda\rightarrow\infty$ if $c\geq 2$. This UV-divergence cannot be eliminated by a counterterm in the Lagrangian due to the high non-locality in the fields of the contribution \eqref{hnl}.

As is well-known, the $S$-matrix is not unitary if time is a noncommutative coordinate \cite{Gomis:2000zz,Seiberg:2000gc,Chaichian:2000ia}. Therefore, for an odd number of spacelike coordinates, unitarity imposes that at least two coordinates are commutative, i.e.\ $c\geq 2$. In this case, as we have seen, some one-loop non-planar diagrams generate divergencies that cannot be regularized by a redefinition of the parameters of the theory. In consequence, as has been shown by V.\ Gayral et al.\ \cite{Gayral:2004cu}, a real scalar field in four dimensions with a quartic self-interaction does not have a well-defined effective action.

On the other hand, notice that the function $\Sigma_{NP}(\hat{p})$ behaves as
\begin{equation}\label{d1}
    \Sigma_{NP}(\hat{p})=\left[(d/2-2)!(m|\Xi\hat{p}|)^{2-d}+\frac{2(-1)^{d/2}}{(d/2-1)!}\log{(m|\Xi\hat{p}|)}\right]
    (1+O(|\hat{p}|^2))\,,
\end{equation}
for small $|\hat{p}|$ and $d>2$. The corresponding result for $d=2$ reads
\begin{equation}\label{d2}
    \Sigma_{NP}(\hat{p})=-2\left[\log{m|\Xi\hat{p}|+\gamma}\right]
    (1+O(|\hat{p}|^2))\,,
\end{equation}
where $\gamma$ is Euler's constant. The result of expression \eqref{d1} evaluated at $d=4$ and $c=0$ corresponds to the contribution to the effective action computed by S.\ Minwalla et al.\ \cite{Minwalla:1999px} by considering one-loop non-planar diagrams.

The divergent behavior at small $\hat{p}$ shown in expressions \eqref{d1} and \eqref{d2} implies that the propagator receives one-loop corrections which are divergent for small values of the momentum in the noncommutative directions. This is the well-known UV/IR-mixing, which shows that in some noncommutative theories the integration of internal momenta can generate divergencies at small values of the external momenta, even for massive fields. As a consequence, these noncommutative theories are non-renormalizable.

\section{Conclusions}\label{conclu}

We have determined the phase space propagator and the phase space generating functional and written a path integral formulation for non-local operators which are relevant in field theories on noncommutative spacetimes. In this formulation, we have derived a master formula (cfr.~\eqref{mf}) for the heat-trace expansion which can be applied to different noncommutative settings. In particular, we considered a non-local operator involving the product of left-Moyal and right-Moyal multiplications.

We have shown that the natural rescaling $x\rightarrow \sqrt{\beta}x$ and $p\rightarrow p/\sqrt{\beta}$ in phase space introduces $O(1/\beta)$ differential operators which act on the potentials. These operators are given by $D_i$, defined in eqs.\ (\ref{df}). Notice that when the heat operator involves only left-Moyal multiplication ($\partial^r=0$ and $\bar{\Theta}=\Theta$) or only right-Moyal multiplication ($\partial^l=0$ and $\bar{\Theta}=-\Theta$) the operators $D_i$ vanish. However, in the case where both left- and right-Moyal multiplications are present, the operators $D_i$ generate SDW coefficients which are non-local even in the generalized Moyal sense. These non-local SDW coefficients are equivalent to the non-planar contributions in a perturbative calculation of the effective action in terms of Feynman diagrams.

Our phase space formulation provides a simple derivation of these results and is suitable for further generalizations in noncommutative models. In particular, the introduction of a Grosse-Wulkenhaar term \cite{Grosse:2003nw,Grosse:2004yu} can be straightforwardly implemented in our phase space approach by replacing the vanishing matrix element of the propagator given by eq.\ \eqref{propagator} by a constant matrix element, proportional to the squared frequency of the harmonic oscillator term. Finally, we consider that this formalism could be a useful tool in the study of other models involving more general non-local operators. Research along these lines is currently in progress.

\section*{Acknowledgments}

The authors thank F.\ Bastianelli for help and suggestions and for participating in the earlier stages of this project. P.A.G.P.\ and S.A.F.V.\ acknowledge D.V.\ Vassilevich for calling their attention on the results of \cite{Gayral:2006vd} and for a discussion on the open problems of heat-trace expansion on the NC torus. O.C.  thanks the Dipartimento di Fisica and INFN Bologna for hospitality and support while parts of this work were completed. The work of O.C.\  is partly funded by SEP-PROMEP/103.5/11/6653. The work of P.A.G.P.\ and S.A.F.V.\ is partly funded by CONICET (PIP 01787) and UNLP (proj. 11/X492). S.A.F.V.'s stay at the Universit\`a di Roma ``La Sapienza'' was partly financed by the ERASMUS MUNDUS Action 2 programme.

\end{document}